\documentclass{article}
\usepackage{booktabs}
\usepackage[it,IT,hang]{subfigure}
\usepackage{hyperref}
\usepackage[margin=1in]{geometry}
\usepackage{authblk}
\usepackage{graphicx}
\usepackage{amsmath}

\title{Eye Gaze Metrics and Analysis of AOI for Indexing Working Memory towards Predicting ADHD}
\author[1]{Gavindya Jayawardena}
\author[2]{Anne Michalek}
\author[1]{Sampath Jayarathna}
\affil[1]{Department of Computer Science}
\affil[2]{Department of Communication Disorders \& Special Education}
\affil[ ]{Old Dominion University, Norfolk, VA 23529}
\affil[ ]{\texttt{gavindya@cs.odu.edu, aperrott@odu.edu, sampath@cs.odu.edu}}
\date{}

\begin{document}
\maketitle

\textbf{Abstract}: ADHD is being recognized as a diagnosis which persists into adulthood impacting economic, occupational, and educational outcomes. There is an increased need to accurately diagnose and recommend interventions for this population. One consideration is the development and implementation of reliable and valid outcome measures which reflect core diagnostic criteria. For example, adults with ADHD have reduced working memory capacity when compared to their peers (Michalek et al., 2014). A reduction in working memory capacity indicates attentional control deficits which align with many symptoms outlined on behavioral checklists used to diagnose ADHD. Using computational methods, such as eye tracking technology, to generate a relationship between ADHD and measures of working memory capacity would be useful to advancing our understanding and treatment of the diagnosis in adults. This chapter will outline a feasibility study in which eye tracking was used to measure eye gaze metrics during a working memory capacity task for adults with and without ADHD and machine learning algorithms were applied to generate a feature set unique to the ADHD diagnosis. The chapter will summarize the purpose, methods, results, and impact of this study.


\section{Introduction}

Attention-Deficit/Hyperactivity Disorder is being recognized as a diagnosis which persists into adulthood impacting economic, occupational, and educational outcomes. Estimates indicate that 3-5\% of adults have a diagnosis of ADHD \cite{willcutt2012prevalence} with prevalence estimated to have increased from 6.1\% of the United States population in 1997 to 10.2\% of the population in 2016 \cite{xu2018twenty}. The disorder is behaviorally marked by difficulty with attention to important details, difficulty initiating and completing tasks, and difficulty modulating behaviors appropriately in relation to the situation \cite{fields2017adult,fostick2017effect}. According to Barkley \cite{barkley1997behavioral}, adult ADHD symptoms result from impairments of inhibition or the inability to regulate and modulate prepotent responses. While a diagnosis of adult ADHD presumes disinhibition, little is known about the physiological underpinnings of that cognitive skill in relation to an adult ADHD diagnosis. There is an increased need to accurately diagnose ADHD through the development and implementation of objective and reliable outcome measures which reflect core diagnostic criteria, like inhibition.  

Researchers in cognitive psychology evidence attention control as the measurable psychological construct which facilitates inhibitory responses by allocating attention according to task demands, especially in the presence of distracting stimuli \cite{conway2005working,engle2002working,kane2001controlled}. Attention control differentiates success during tasks requiring intentional and sustained constraints for effective inhibition, like dichotic listening \cite{colflesh2007individual} or processing speech in noise \cite{ronnberg2013ease}. Measurements of attention control are demonstrated through differences in working memory capacity (WMC) accounting for approximately 60\% of the variance seen across people on measures of WMC, like complex span tasks \cite{engle1999working}. Adults with ADHD have reduced WMC when compared to their peers \cite{michalek2014effects} and, despite the understanding that disinhibition is central to an ADHD diagnosis and differences in WMC mathematically represent the resource which makes inhibition possible, there is a paucity of research investigating physiological responses during measures of WMC which could differentiate adults with and without ADHD. 

The two goals of this work is to determine the feasibility of identifying and integrating eye gaze metrics from a WMC task using machine learning to generate a valid and reliable feature set which indexes and predicts an ADHD diagnosis and understand the utility of area of interest (AOI) in order to capture eye gaze metrics and predict ADHD in the context of a WMC task using machine learning algorithms.
This chapter investigates gaze measures that map onto these valid neurocognitive deficits that are central to ADHD within the context of a WMC task.
In addition, this chapter investigates the differences between ADHD and non-ADHD participants by analyzing two main sequence relationships of saccade amplitude towards saccade duration and peak velocity.
The development of these objective measures of ADHD  will facilitate its diagnosis and reveal strategies that can enhance the future design effective intervention strategies and accessible classroom environments.

\section{Background}

\subsection{Working Memory Capacity (WMC) Tasks}

Working memory is the cognitive system which makes it possible to mentally hold and manipulate information simultaneously. Over a decade of work by Engle and colleagues \cite{conway2008variation} supports the use of complex span tasks as not only measures of working memory but as a reflection of individual differences in WMC. Performance on complex span tasks generates a composite working memory score numerically indicating how well someone can manipulate and hold information. However, differences in WMC or that composite score represent a person’s ability to moderate and control attention. Adults with ADHD have reduced working memory when compared to their peers demonstrating significant differences in WMC \cite{alderson2013attention}. This finding suggests that adults with ADHD have a reduced ability to monitor and control attention, especially during situations with competing stimuli \cite{michalek2014effects} or that require response inhibition \cite{lee2015saccadic,roberts2011separating}. Little is known about the underlying covert processes engaged during inhibitory tasks which rely on attention allocation. Using physiological measures during a task which validly reflects attention control, like a complex span task, provides objective diagnostic information for adults with ADHD.

The reading span task (R-Span) is one complex span task widely used as a valid measure of working memory yielding a WMC score \cite{conway2005working}. The R-span was originally developed by Daneman and Carpenter (1983) \cite{daneman1983individual} as a predictor of reading comprehension and was subsequently modified by Engle and colleagues \cite{engle1999working}. During the R-Span people see one sentence on a computer screen, read the sentence out loud, determine if the sentence is meaningful with a yes or no response, and then verbally identify a letter typed at the end of each sentence. After a set of sentences, the person is asked to verbally recall as many letters as possible in order of presentation. This task represents the person’s ability to hold and manipulate information simultaneously. To date, there have been no empirical studies investigating eye gaze metrics collected during this task which might differentiate performance and further explain diagnostic differences for adults with and without ADHD. 

\subsection{Machine Learning and ADHD}

Typically, experts diagnose ADHD using subjective checklists and related academic and cognitive performance measures \cite{greenhill1998diagnosing}.
These comprehensive assessments can be time consuming, inconsistent, and can inaccurately represent deficits making differential diagnosis challenging. Ideally, it would be more efficient and reliable to develop predictive algorithms based on physiological metrics reflecting core diagnostic criteria. However, designing this type of computer program can be difficult because there is not an existing set of confirmed mathematical features which accurately differentiate between adults with and without ADHD. Machine learning principles offer a solution to this barrier. While it is not practical to develop algorithms by providing a specific set of instructions, machine learning uses numeric features representing a cognitive skill to teach the computer data patterns and inferences which can be applied to groups of data for predicting accurate classifications. 

In machine learning there are several varieties of subcategorical learning algorithms. Supervised learning is an example of such a subcategory. The core core objective of supervised learning is to build a mathematical model which can be used to predict the outputs of new samples using the training data. Usually, the training data set is stored in a matrix. Each row of the training data matrix corresponds to one training instance which also contains the desired output. For example, in the ADHD/Non-ADHD literature par supervised learning algorithms are iterated through the training dataset to learn a mathematical model to predict or classify the output associated with unseen inputs. When determining whether a person does or does not have ADHD using eye gaze as the outcome metric, the training data would consist of eye tracking data of each person and each person would have a class specifying whether that person is identified as having ADHD or not having ADHD. Supervised learning algorithm will build a general mathematical model which covers the training data space. When a previously unseen person's eye gaze data is entered, the model will use its past experience to accurately predict whether or not that person has ADHD.

Some experts may question the reliability and validity of using machine learning for predicted outputs because those outputs do not consider experts from psychology or medicine in the decision process. Even with the involvement of expert physicians, it is reported that “diagnostic errors contribute to approximately 10 percent of patient deaths”, by Institute of Medicine at the National Academies of Science, Engineering and Medicine\cite{national2016improving}. Causes for such diagnostic errors could be communication errors between patients and physicians and other failures of the healthcare system. These challenges could be addressed by identifying patterns of the symptoms patients confirm and use them to predict potential diagnostic codes. Even if there is a lack of communication between the patient and the physician or there is a failure in a healthcare system, the symptom patterns of the patient would be highlighted so that accurate diagnosis prediction could be facilitated. 

Currently, machine learning is being used for diagnostic prediction not only based on reported symptoms, but also based on patient history and data extracted from wearable devices. Classification algorithms compare the symptoms pattern of the patient and other related data with the other patients in the training dataset in order to predict an accurate diagnosis. In literature using machine learning for ADHD diagnosis and classification, the first attempt to classify adult ADHD patients and healthy controls using a machine learning algorithm is \cite{mueller2010classification}. They conducted research to classify ADHD patients and healthy controls using support vector machine (SVM) learning based on event related potential (ERP) components. They examined data from 148 adult participants. Among them, 50\% \cite{mueller2010classification} were diagnosed as ADHD while the rest did not have a diagnosis of ADHD. Both groups of adults were selected in a manner that age and gender did not vary between the two groups \cite{mueller2010classification}. Each participant performed a visual two stimulus GO/NOGO task \cite{mueller2010classification} and ERP responses of participants were decomposed into independent components and created the feature set. Classification accuracy of assigning ADHD participants and healthy controls to the corresponding groups using a non-linear SVM with 10-fold cross-validation was 92\% \cite{mueller2010classification}, whereas it was 90\% \cite{mueller2010classification} for linear SVM. This research suggests that classification by means of non-linear methods is more accurate for experiments conducted in a clinical context \cite{mueller2010classification}. Even-though this study uses machine learning approaches to predict ADHD, it does not use eye gaze metrics nor working memory capacity.

\cite{peng2013extreme} shows that extreme learning machine (ELM), a machine learning algorithm, achieves 90.18\% \cite{peng2013extreme} accuracy when predicting ADHD using structural MRI data. This study confirmed that linear support vector machine and support vector machine-RBF achieves an accuracy of 84.73\% and 86.55\% \cite{peng2013extreme} respectively when the same structural MRI dataset is used. Both extreme learning machine and support vector machine have been evaluated to find the classification accuracy using cross-validation. The goal of their study was proposing an ADHD classification model using the extreme learning machine (ELM) algorithm for ADHD diagnosis. They assessed the computational efficiency and the effect of sample size on both extreme learning machine and support vector machine. They acquired MRI images from 110 participants with 50\% of them having a diagnosis of ADHD \cite{peng2013extreme}. This study gives us insight about how applying a machine learning model can accurately predict ADHD. 

Moreover, \cite{marcano2016classification} aimed to classify people with ADHD and  wihtout ADHD using autoregressive models. They used EEG data collected using 26 electrodes from a group of children between the ages of 6 and 8 \cite{marcano2016classification} to discriminate between ADHD and Non-ADHD. Children participated in multiple experimental conditions, such as eyes open, eyes closed, and quiet video baseline tasks while collecting EEG data \cite{marcano2016classification}. This study ferified that KNN classifier is able to provide high classification accuracy when classifying children as either ADHD or typically delveoping ADHD. The accuracy achieved in this study was high and varying between 85\% and 95\% \cite{marcano2016classification}.

Finally, \cite{abibullaev2012decision} used EEG data with semi-supervised learning in order to predict a ADHD and Non-ADHD diagnosis. They had 10 children participants with 7 of them having a diagnosis of ADHD and 3 were typically developing \cite{abibullaev2012decision}. They trained and tested support vector machine with EEG data of each participant achieving an accuracy of 97\% \cite{abibullaev2012decision} for ADHD prediction using support vector machine learning. 

Taken together, the literature affirms the successful use of machine learning to accurately predicate a diagnosis of ADHD. However, an empirical gap exists with regard to the training dataset used to train the machine learning model. The majority of studies conducted have primarily used MRI, fMRI, or EEG data to train the machine learning model used to discriminate between ADHD and Non-ADHD. One of our goal of this project is to predict diagnosis of ADHD using eye gaze metrics and measures of working memory as the training data for the machine learning algorithms. 

In this project we predicted a diagnosis of ADHD using eye gaze metrics collected during a WMC task. The output of our task is limited to two classes, ADHD and Non-ADHD. We used supervised learning classification algorithms and we evaluated the predicted output in terms of how close it is to the actual output. We evaluated outcomes using an accuracy, precision, recall, and f-measure evaluation metrics. Together these metrics gave us an good understanding about the performance of the classifier.

\subsection{Area of Interests and Attention Tasks}

The studies \cite{mohammadhasani2015link} and \cite{serrano2018children} have considered AOIs of stimuli for statistical analysis of the eye tracking. The study \cite{kaiser2019biased} compared borderline personality disorder (BPD) patients to Cluster-C personality disorder (CC) patients and non-patients (NP) regarding emotion recognition in ambiguous faces and their visual attention allocation to the eyes which is the AOI. The authors have found BPD have a  biased visual attention towards the eyes. \cite{fong2019immediate} is a another study which investigates the immediate effects of coloured overlays on reading performance of preschool children with ASD. The authors of the study have used eye tracking and concluded that coloured overlays may not be useful to improve reading and ocular performance in children with ASD in a single occasion \cite{fong2019immediate}.

According to the literature, AOIs have been used in multiple studies related to various attention tasks. Even though AOIs have been used, there is a paucity of using eye movement fixations and saccades occurred within the AOIs when completing a WMC measure for ADHD classification using machine learning approach.

We look at the consistency and stability of eye movement fixations and saccades occurred within the AOIs of stimuli when completing a WMC measure. This is an important line of inquiry because it investigates how relevance may be reflected in eye movements features for atypical and complex attentional systems, such as in the context of ADHD. 

\subsection{Eye Movements and ADHD}
Eye movement behavior is a result of complex cognitive processes; therefore, eye gaze metrics can reveal objective and quantifiable information about the quality, predictability, and consistency of these covert processes \cite{van2007oculomotor}. Eye gaze measurement includes a number of metrics relevant to oculomotor control \cite{komogortsev20132d} including saccadic trajectories, fixations, and other relevant measures - such as velocity, duration, amplitude, pupil dilation \cite{krejtz2018eye}. A saccade (rapid eye movement from one fixation point to another) itself may not be an informative indicator of cognition since visual perception is suppressed during a saccade. However, fixations require preceding saccades to help place the gaze on target stimuli to gather salient and relevant information. We believe that analysis of these eye movements can provide important cumulative clues about the underlying physiological functions of attention control during a WMC task which can differentiate a diagnosis of ADHD for adults.

There is substantial overlap in brain systems that are involved in oculomotor control and cognitive dysfunction in ADHD. The precise measurements of eye movements during cognitively demanding tasks provide a window into underlying brain systems affected by ADHD. The neural substrates of oculomotor control are well established \cite{leigh2004using} and show proximity to and overlap with the cortical and subcortical structures involved in cognitive dysfunction in ADHD. For example, the cortical structures that mediate saccadic programming as well as a number of saccadic behaviors include frontal-parietal areas such the frontal eye field (FEF), supplementary eye field (SEF), parietal eye field (PEF), and DLPC. These areas are also affected during cognitive control and WM in ADHD \cite{rubia2018cognitive}. With respect to subcortical structures, the accuracy of saccades is maintained via cerebellum. For example, saccadic hypometria is an undershooting of a saccade to a target that is typically seen in normal subjects, whereas saccadic hypermetria, overshooting the target, is a hallmark feature of cerebellar dysfunction \cite{leigh2015neurology}. A study of saccades during visuo-spatial WM has reported significant diagnostic group differences in under- versus over-shooting to the target between boys with ADHD and non-affected controls, such that the ADHD group tended to overshoot the target and the control group tended undershoot the target \cite{rommelse2008deficits}. Studies have shown that individuals with ADHD also have deficits in the suppression of saccades relative to controls \cite{mostofsky2001oculomotor,munoz2003altered,rommelse2008deficits}. Similarly, people with ADHD demonstrate difficulties with intentionally inhibiting ocular responses when compared to their peers during tasks which require purposeful anti-saccade behaviors \cite{lee2015saccadic,roberts2011separating}. Eye gaze metrics, especially saccade features, reliably reveal important differences between adults with and without ADHD.

Based on the diagnostic utility of eye movements, \cite{blazey2003adhd} invented a method which determines whether an individual has ADHD by sampling the eye movements of participants when they are in an inactive state. Patented as "ADHD detection by eye saccades" \cite{blazey2003adhd}, their  procedure includes a sampling device which has infrared radiation for brightening the eye of a participant and detecting reflections from the eye. The eye movement data collected using their device determines the value of a pre-selected parameter which has a threshold value indicating whether the participant has ADHD or not. According to their study, the most significant feature of the eye movement data is the angular acceleration of the eyeball \cite{blazey2003adhd}. They have measured ocular angular acceleration for the participants by asking them to stare at a blank screen \cite{blazey2003adhd}. The measurement data of the angular acceleration of the eye below the threshold value indicates diagnosis of ADHD and the data above the threshold value constitutes a classification of healthy/normal. Although the study conducted by \cite{blazey2003adhd} used on a mechanical device with infrared radiation and not machine learning, the fact that it acquired a patent supports its strong impact and confirms that eye measurements could be used to diagnose ADHD. 

\subsection{Eye Gaze and Machine Learning and ADHD}

There is a paucity of empirical studies which implement machine learning to predict ADHD classification using a measures of attention control or WMC. However, there are a few investigations which use maching learning in combination with measures of inhibition. \cite{hart2014pattern} measured activation patterns using functional magnetic resonance imaging while adolescents with ADHD performed a Stop Task. During this task, participants had to suppress or inhibit the motoric response of pushing a button. The researchers used Gaussian process classifiers and whole activation pattern analysis and were able to predict the ADHD diagnosis with 77\% accuracy \cite{hart2014pattern}. Likewise, in a study with adults with and without a diagnosis of ADHD, machine learning predicted the diagnosis with a specificity of .91 and sensitivity of .76 based on EEG metrics during a NoGo task measuring inhibition \cite{biederman2017diagnostic}. These results support collecting physiological metrics during tasks required attention control to generate pattern recognition analysis for the accurate classification of ADHD. 

Similarly, only one study was located which used eye movements in conjunction with machine learning to predict ADHD \cite{tseng2013high}. This study included participants diagnosed with ADHD, fetal alcohol spectrum disorder (FASD), and Parkinson's disease (PD). Researchers presented short video clips to each participant and analyzed the resulting data sets for three specific types of eye movement features: 1) oculomotor-based features such as fixation durations and distributions of saccade amplitudes; 2) saliency-based features; and 3) group-based features \cite{tseng2013high}. Results confirmed that saliency based features best differentiated children with ADHD and FASD from typically developing children. Machine learning algorithms predicted ADHD in the sample of children with 77.3\% accuracy \cite{tseng2013high}. Taken together, these empirical findings suggest that diagnostic biomarkers of ADHD could be generated from eye gaze metrics during a WMC task using machine learning. As such, in this feasibility study, we examined patterns of saccades and stability of fixations generated when completing a measure of WMC to create a feature set which could be used to differentiate a diagnosis of ADHD for adults. Based on the evidence that WMC is reduced in adults with ADHD \cite{michalek2014effects}, measurement of eye movements during a measure of WMC will address the following research question:  1) do eye gaze feature values indexing a WMC task predict the classification of ADHD in adults? 

\section{Methodology}

\subsection{Participants}
A total of 14 adult participants without (n = 7) and with a diagnosis (n = 7) between the ages of 18-35 were recruited for this study from an higher education institution in the mid-Southeastern United States. The seven adult participants were (6 F, 1 M, M\textunderscore{age}=22.85, SD\textunderscore{age}=3.01) diagnosed with ADHD by medical practitioners and that diagnsosis was confirmed through formal and verified documentation. Each ADHD participant also completed an informational interview verbally confirming their diagnosis. Moreover, adults with ADHD remained medication free for the 12 hours prior to study participation. Prior to beginning study tasks, all adults were informed of their risks regarding remaining medication free and participating in the study. Participants provided their consent by signing forms outlining costs and benefits of participation approved the University's Institutional Review Board (IRB) in accordance with the Helsinki Declaration. Participants who completed the protocol were given a ten dollar Amazon or Chick-Fil-A gift card. 

\subsection{Working Memory Capacity Task}
WMC is reflected through complex span tasks, including the Reading Span (R-Span). The R-Span is a validated task designed to reflect the cognitive system's ability to maintain activated representations \cite{engle1999working,engle2002working}. In the R-Span task, participants are asked to read a sentence and letter they see on a computer screen. Sentences are presented in varying sets of 2-5 sentences. Participants are asked to judge sentence coherency by saying 'yes' or 'no' at the end of each sentence. Then, participants are asked to remember the letter printed at the end of the sentence. After a 2-5 sentence set, participants are asked to recall all the letters they can remember from that set. WMC scores are generated based on the number of letters accurately recalled divided by the total number of possible letters recalled in order. However, this project focused on measures of visual attention which could differentiate adults with and without ADHD. 

\subsection{Apparatus}
Eye gaze metrics were recorded and analyzed using the Tobii Pro X2-60 computer screen-based eye tracker with Tobii Studio analysis software. The Tobii Pro X2-60 records eye movements using infrared corneal reflective technology at a sampling rate of 60 Hz (i.e. approximately once every 16.23 milliseconds). Gaze data accuracy was within 0.4 degrees of visual angle and precision was within 0.34 degrees of visual angle. Tobii's eye tracking technology is effective for generating reliable and valid brain/behavior outcomes for children and adolescents \cite{richmond2009relational}.

All of the participants fulfilled the following inclusion criteria:
1) between 18 and 65 years of age, 2) spoke English as their first language, 3) self-reported normal vision with or without corrective lenses, 4) no history of psychotic symptoms; and 5) no comorbid cognitive impairments (e.g. documented learning disabilities, reading disabilities).

\begin{figure*}[hbt!]
  \centering
  \includegraphics[width=\linewidth]{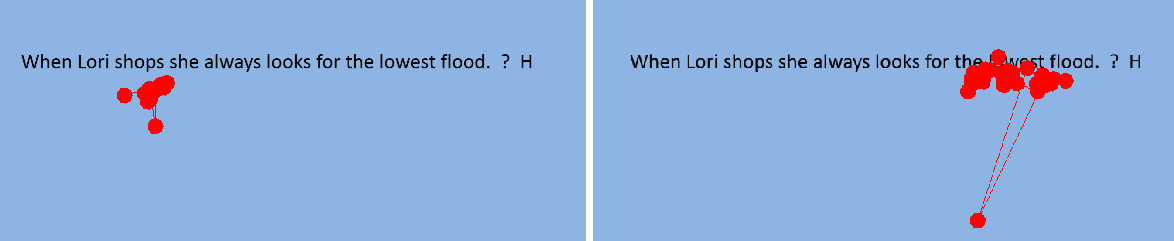}
  \caption{Comparison of Eye Fixatoins for ADHD (Left) and Non-ADHD (Right) particiapnt during WMC Task.}
  \label{fig:rspan}
\end{figure*}

\subsection{Eye Movement Features}
The human oculomotor plant (OP)\cite{komogortsev2010biometric} consists of the eye globe and six extraocular muscles and its surrounding tissues, ligaments each containing thick and thin filaments, tendon-like components and liquids. In general there are six major eye movement types: fixations, saccades, smooth pursuits, optokinetic reflex, vestibule-ocular reflex and vergence\cite{leigh2015neurology}. An eye-tracker provides eye gaze position information as well as other gaze related parameters (pupil dilation etc.) so that algorithmic derivation in terms of two primary eye movements, fixations (relative gaze position at one point on the screen) and saccades (rapid eye movements of gaze from one fixation point to another) can be analyzed to derive the users attention patterns.

We are interested in investigating number of eye fixation based features in the current framework. We developed a detailed saccade and fixation feature set using the following qualifiers: gender, number of fixations, fixation duration measured in milliseconds, average fixation duration in milliseconds, fixation standard deviation in milliseconds, pupil diameter left, pupil diameter right, and diagnosis label or class. Due to the sampling rate of the tracking system, we were not able to calculate microsaccades and overshoot/undershoot saccades as components of the feature set.

\subsection{Measuring Attention during WMC}

The data for this study was collected during a larger project involving adults with and without ADHD and an audiovisual listening in noise task where WMC scores were measured and used as a cognitive covariate without eye tracking metrics. The entire testing session for the project took approximately 45 minutes. The session began with the participant interview, explanation of the purpose of the study, and review of the consent form. During the interview, the participants provided demographic information and were screened to confirm that all inclusion criteria were satisfied. Once participants indicated they understood their rights and gave consent, they entered the testing area to begin the study. Participants sat at a desk in front of a Dell Computer with a 21 inch monitor. The distance and position of each participant was modified in order to maintain a 45 degree viewing angel of the monitor. For each participant, the experimental tasks began with eye gaze calibration. Once calibration was confirmed, participants viewed a welcome screen followed by the random presentation of several experimental tasks, including the RSPAN task. The location of the RSPAN in the order of experimental tasks was randomized and counterbalanced across participants in order to maintain validity. Participants were randomly assigned to a group determining the order of experimental task presentation prior to beginning the study. For all of the experimental tasks, participants were given practice trials. 

\section{Machine Learning on Data}
We chose precision, recall, f-measure, and accuracy as the evaluation measures for our work. Prior studies \cite{manevitz2007one} have already proven that these measures are independent of category distributions provided that precision and recall are measured at the same time. Intuitively, precision measures exactness of the system (i.e., out of all predicted data instances for a specific category label how many are predicted correctly) while recall indicates the completeness of the system (i.e., out of all labeled data for a specific a category label how many are predicted correctly). F value measures the balance between precision and recall in a single value. In our tables with results assessing classifiers, precision, and recall refers to their weighted average values. Accuracy specifies the fraction of the predictions that the classifier predicated correctly. We employed a grid search mechanism to identify the best parameter combination for optimal result. The optimal parameters are selected based on performance for each classifier after a 10-fold cross validation. 

\begin{table}[hbt!]
\centering
  \caption{RSPAN Score of the ADHD Vs Non-ADHD}
  \label{tab:rspan}
  \begin{tabular}{ccccl}
    \toprule
    Participant & Age & Gender & RSPAN & Classification\\
    \midrule
    3 & 18 & Female & 0.86 & Non-ADHD\\
    7 & 35 & Male & 0.88 & Non-ADHD\\
    9 & 19 & Female & 0.60 & Non-ADHD\\
    17 & 23 & Male & 0.55 & Non-ADHD\\
    20 & 21 & Female & 0.57 & Non-ADHD\\
    25 & 32 & Male & 0.88 & Non-ADHD\\
    26 & 20 & Female & 0.74 & Non-ADHD\\
    \midrule
    30 & 21 & Female & 0.51 & ADHD\\
    34 & 19 & Male & 0.67 & ADHD\\
    35 & 26 & Female & 0.76 & ADHD\\
    36 & 29 & Female & 0.71 & ADHD\\
    37 & 21 & Female & 0.60 & ADHD\\
    38 & 21 & Female & 0.40 & ADHD\\
    47 & 23 & Female & 0.62 & ADHD\\
  \bottomrule
\end{tabular}
\end{table}

Table~\ref{tab:rspan} shows the RSPAN score for the participants in the current study. An independent t-test statistical analysis (p=0.07) confirms that for this feasibility study there are no significant group differences on WMC scores and are predicted to be a result of the small sample size. The RSPAN is an individual differences measure and significance in variance is detected with large sample sizes. Additionally, WMC scores are typically generated through a composite score of two or more span tasks \cite{conway2005working}, for example, a previous investigation by one of the authors confirmed group differences in WMC using the RSPAN and operation span (OSPAN) to generate a WMC composite score for adults with ADHD \cite{michalek2014effects}.

\subsection{Visual Analysis}

Figure ~\ref{fig:rspan} presents images of eye gaze patterns from two adults participants, one with and one without ADHD. Informal visual analysis indicates that the adult with ADHD is fixating primarily below the stimulus items with little direct fixation to sentence components including: the words, the decision point, or the item to be remembered. Unlike the adults with ADHD, the adult without ADHD has a majority of fixations which are in-line with all sentence components. Although this is a conclusion generated from informal visual inspection, it reveals that adults with ADHD are not visually scanning stimulus items in a path similar to adults without ADHD. The fixation cluster pattern is just below the stimulus sentence components. This is consistent with the findings of Krejtz et al. (2015) who suggest that while adults with ADHD had similar fixations to salient visual cues when compared to adults without ADHD, they demonstrated less structured and more chaotic scan patterns \cite{krejtz2015gaze}. 

\subsection{Machine Learning for Classification Prediction}

We generated three feature sets categorized according to metric type: 1) fixation feature set; 2) saccade feature set; and 3) saccade and fixation combination feature set. Each of the three feature sets were individually entered into 43 different classifiers yielding precision rates, recall rates, F1 scores, and percent accuracy. We identified six of the top performing classifiers for each of the three feature sets: J48, LMT, RandomForest, REPTree, K Star, and Bagging. Results for each feature set are discussed individually.

Six of the top performing classifiers for the fixation feature set are listed in the Table ~\ref{tab:classifiers}. The Bagging classifier (ensemble meta-estimator) yielded the highest percent accuracy with 78.48\% indicating that a fixation feature set collected during a RSPAN task classifies a diagnosis of ADHD with greater than 70\% accuracy. The REPTree classifier yielded the lowest percent accuracy at 76.77\%. 

\begin{table}[hbt!]
\centering
  \caption{Classification of Eye Fixation Features during WMC}
  \label{tab:classifiers}
  \begin{tabular}{ccccl}
    \toprule
    Classifier & Precision & Recall & F1 & Accuracy\\
    \midrule
    J48 & 0.77 & 0.76 & 0.77 & \textbf{77.79} \\
    LMT & 0.77 & 0.77 & 0.77 & \textbf{77.92} \\
    RandomForest & 0.76 & 0.76 & 0.76 & 76.79 \\
    REPTree & 0.75 & 0.76 & 0.75 & 76.77 \\
    K* & 0.76 & 0.76 & 0.76 & 76.92 \\
    Bagging & 0.77 & 0.78 & 0.77 & \textbf{78.48} \\
  \bottomrule
\end{tabular}
\end{table}

\begin{figure}[hbt!]
  \centering
  \includegraphics[trim={0pt 0pt 0pt 0pt},clip,height=.400\textheight]{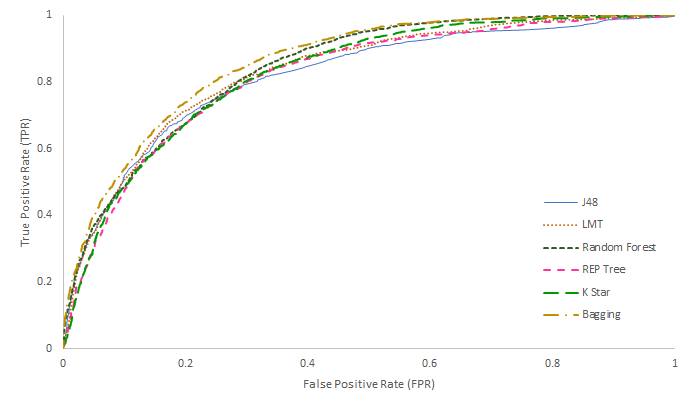}
  \caption{ROC Graph of the Top Performing Classifiers for Fixation Feature Set.}
  \label{fig:roc_fixations}
\end{figure}

To further investigate the performance metrics for the 6 most effective classifiers for the fixation feature set we  generated a Receiver Operating Characteristics (ROC) graph (see Figure ~\ref{fig:roc_fixations}). The ROC graph displays the relative trade-off between benefits (true positive) rates on the Y axis and the costs (false positive) rate on the X axis. The graph shows the Bagging as our top performing classifier offering the best trade-off in terms of the cost and the benefits. 

Table ~\ref{tab:saccade_classifiers} outlines six of the top performing classifiers for the saccade feature set. The Random Forest classifier yielded the highest percent accuracy at 91.14\% indicating that the saccade feature set collected during a RSPAN task classifies a diagnosis of ADHD with greater than 90\% accuracy. The J48 classifier yielded the lowest percent accuracy at 88.95\%.

\begin{table}[hbt]
\centering
  \caption{Classification of Saccade Features during WMC}
  \label{tab:saccade_classifiers}
  \begin{tabular}{ccccl}
    \toprule
    Classifier & Precision & Recall & F1 & Accuracy\\
    \midrule
    J48 & 0.89 & 0.89 & 0.89 & 88.95 \\
    LMT &  0.89 &  0.89 & 0.89 & \textbf{89.51} \\
    RandomForest & 0.91 & 0.91 &   0.91 & \textbf{91.14} \\
    REPTree &  0.89 & 0.89 &  0.89 & 89.16 \\
    K* & 0.86 & 0.86 & 0.86 & 85.98 \\
    Bagging & 0.91 & 0.91 & 0.91 & \textbf{90.82} \\
  \bottomrule
\end{tabular}
\end{table}

\begin{figure}[hbt!]
  \centering
  \includegraphics[trim={0pt 0pt 0pt 0pt},clip,height=.400\textheight,width=.750\linewidth]{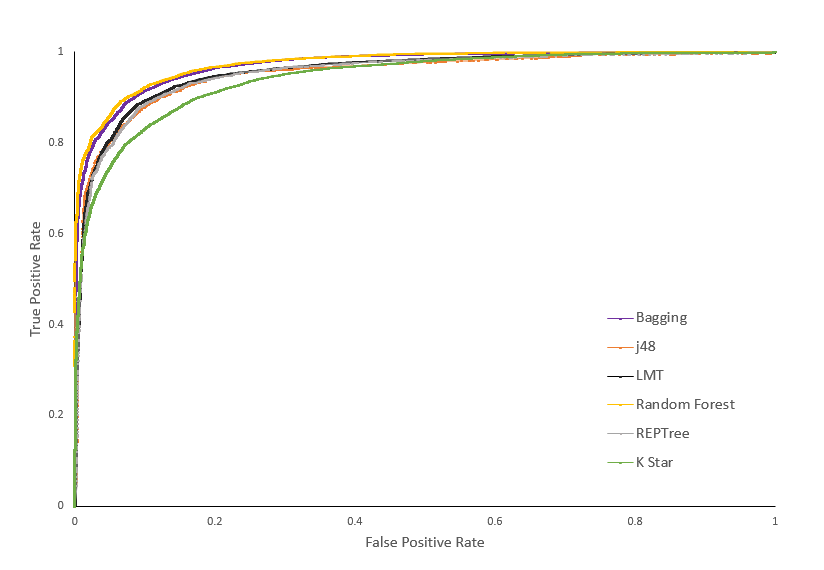}
  \caption{ROC Graph of the Top Performing Classifiers for Saccade Feature Set.}
  \label{fig:roc_saccades}
\end{figure}

We  generated a ROC graph (see Figure ~\ref{fig:roc_saccades}) for the classifiers we selected for the saccade feature set to investigate the performance metrics. The ROC curve shows that Random Forest classifier has the largest Area Under the Curve (AUC) meaning that it has the lowest error. The AUC of Random Forest classifier is 0.9114. Therefore, it has 91.14\% chance of correctly distinguishing between ADHD and Non-ADHD. Random Forest is our top performing classifier offering the best trade-off in terms of the cost and the benefits for the saccade feature set.

Finally, table ~\ref{tab:fixation_and_saccade_classifiers} provides results of the six top performing classifiers for the combination of saccade and fixations feature set. The Random Forest classifier yielded the highest percent accuracy at 91.11\% indicating that the combination of fixation and saccade features collected during a RSPAN task classified a diagnosis of ADHD with greater than 90\%accuracy. The K Star classifier yielded the lowest percent accuracy at 77.21\%. 

\begin{figure}[hbt!]
  \centering
  \includegraphics[trim={0pt 0pt 0pt 0pt},clip,height=.400\textheight]{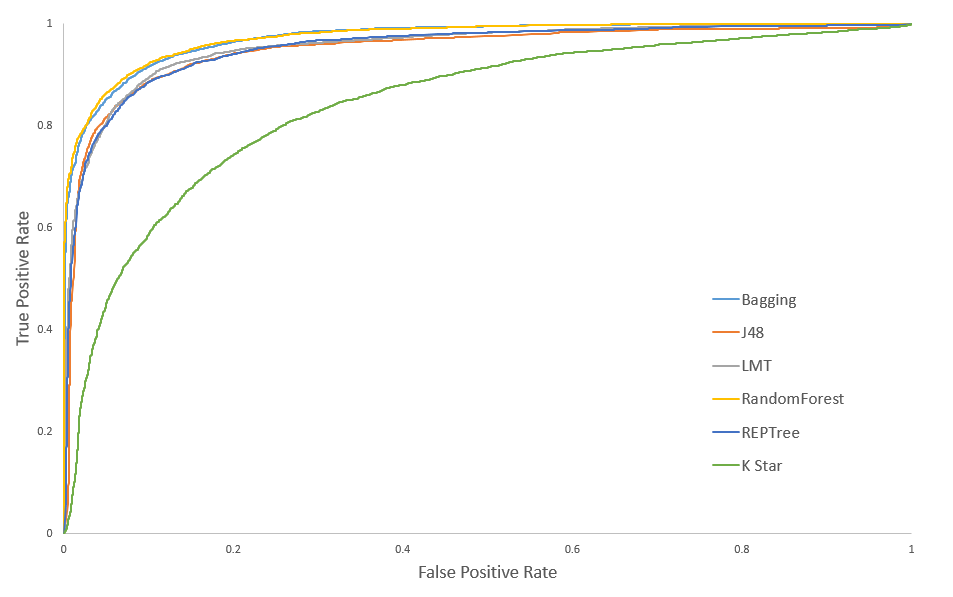}
  \caption{ROC Graph of the Top Performing Classifiers for the Combination of Fixation and Saccade Features Set.}
  \label{fig:roc_all}
\end{figure}

\begin{table}[hbt!]
\centering
  \caption{Classification of Eye Fixation and Saccade Features during WMC}
  \label{tab:fixation_and_saccade_classifiers}
  \begin{tabular}{ccccl}
    \toprule
    Classifier & Precision & Recall & F1 & Accuracy\\
    \midrule
    J48 & 0.89 & 0.89 & 0.89 & 89.19 \\
    LMT & 0.89 & 0.89 & 0.89 & \textbf{89.91} \\
    RandomForest & 0.91 & 0.91 & 0.91 & \textbf{91.11} \\
    REPTree & 0.89 & 0.89 & 0.89 & 89.16 \\
    K* & 0.77 & 0.77 & 0.77 & 77.21 \\
    Bagging & 0.91 & 0.91 & 0.91 & \textbf{90.83} \\
  \bottomrule
\end{tabular}
\end{table}

We generated a ROC graph (see Figure ~\ref{fig:roc_all}) to investigate the performance metrics for the classifiers we selected for the combination of saccade and fixations feature set as well. The graph shows that even for the combination of saccade and fixations feature set, Random Forest is the top most performing classifier offering the best trade-off in terms of the cost and the benefits. The AUC of Random Forest classifier is 0.9111 meaning that it has 91.11\% chance of correctly distinguishing between ADHD and Non-ADHD.

\section{Main-Sequence Relationship}
We investigated the main sequence relationships among saccade amplitude, saccade duration, and saccade peak velocity for both ADHD and non-ADHD groups. Feature sets for the main sequence relationships are based on the following qualifiers: saccade amplitude measured in degrees, saccade duration measured in milliseconds, and saccade peak velocity measured in degrees per second. Saccade amplitude is the size of a saccade. Saccade peak velocity is the highest velocity reached during a saccade. Saccade duration is the time taken to complete the saccade.

Saccade peak velocity is calculated by the Equation. \ref{eq:velocity_amp} in degrees/second \cite{leigh1999neurology}
\begin{equation}
    \label{eq:velocity_amp}
    \dot{\theta}_{peak\_velocity} = \allowbreak \dot{\theta}_{max} \times (1  - e^{-\theta_{amplitude}/C}) 
\end{equation}
where ${\dot{\theta}_{peak\_velocity}}$ is the saccade peak velocity, ${\dot{\theta}_{max}}$ is the asymptotic peak velocity (500 degree/second), $\theta_{amplitude}$ is the saccade amplitude (degrees) and C is the constant (14 for normal humans). Figure \ref{fig:amplitude_vs_velocity_Normal} shows the relationship between saccade amplitude and saccade peak velocity for normal humans.

Saccade duration in milliseconds is calculated by Equation. \ref{eq:duration_amp} \cite{carpenter1988movements}. Figure \ref{fig:amplitude_vs_duration_Normal} shows the relationship between saccade amplitude and saccade duration for normal humans.

\begin{equation}
    \label{eq:duration_amp}
    t_{duration} = (2.2 \times \theta_{amplitude} + 21) 
\end{equation}

\section{Area of Interests}
\begin{figure*}[hbt!]
  \centering
  \includegraphics[width=1.0\linewidth,height=.15\textheight]{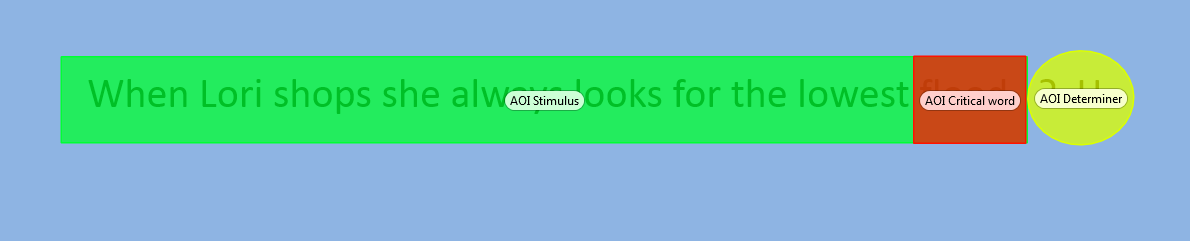}
  \caption{AOIs During WMC Task from a Sentence generated using Tobii Studio Analysis Software.}
  \label{fig:aoi}
\end{figure*}

In addition, we employed a number of fixation and saccade based features captured within three AOI groups in sentences presented in RSPAN task. We utilized a standard RSPAN task where participants are instructed to read a sentence and a letter displayed on a computer screen, judge the sentence's coherency, and memorize the letter at the end. We extracted their eye movement features based on three stimuli: 1) area of the sentence, 2) area of the critical word that determines the coherency of the sentence, and 3) the decision area with the letter to be remembered. Figure. \ref{fig:aoi} shows the three AOIs drawn on a single sentence using Tobii Analysis software. Note that the boundaries of AOIs are drawn manually (static AOIs).

We derived two feature sets for the investigation of fixations and saccades within AOIs based on the following qualifiers: number of fixations in AOI 1, 2 and 3, fixation duration in AOI 1, 2 and 3, average fixation duration in AOI 2, fixation standard deviation in AOI 2, pupil diameter of both eyes in AOI 2 and 3, maximum and minimum saccade amplitude in AOI 1, 2 and 3, average saccade amplitude in AOI 1, 2 and 3, and standard deviation of saccade amplitude in AOI 1, 2 and 3, respectively.

\begin{itemize}
    \item {\em Scene-based}: Feature set including the above qualifiers within the AOIs of sets of 2-5 sentences. 
    \item {\em Sentence-based}: Feature set including the above qualifiers within the AOIs of all the sentences.
\end{itemize}

All fixation features and saccade features were calculated using Pandas, a Python data analysis library. Prior studies \cite{diamantopoulou2005adhd} suggested that diagnostic criteria for ADHD should be adjusted to gender differences. We find that including gender in the feature set slightly increases the performance across all our classifiers.

\section{Analysis of Main Sequence Relationships}

\begin{figure*}[hbt!]
\centering
\subfigure[Amplitude vs. Peak Velocity of ADHD]{%
    \label{fig:amplitude_vs_velocity_ADHD}
    \includegraphics[trim={0pt 0pt 0pt 20pt},clip,height=.400\textheight,width=.25\textwidth]%
        {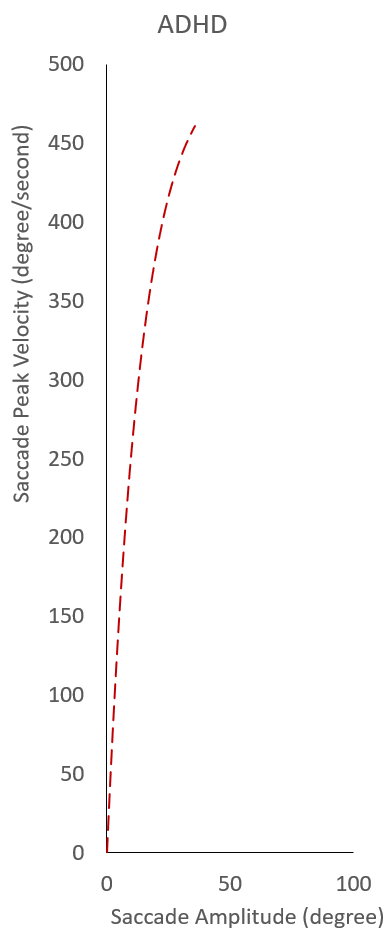}
}
\hspace{1em}%
\subfigure[Amplitude vs. Peak Velocity of Non-ADHD]{%
    \label{fig:amplitude_vs_velocity_Non_ADHD}
    \includegraphics[trim={0pt 0pt 0pt 20pt},clip,height=.400\textheight,width=.25\textwidth]%
        {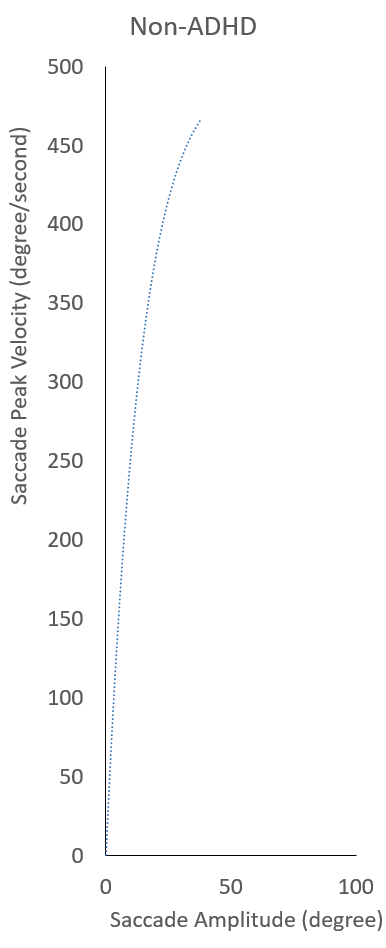}
}
\hspace{1em}%
\subfigure[Amplitude vs. Peak Velocity of Normal]{%
    \label{fig:amplitude_vs_velocity_Normal}
    \includegraphics[trim={0pt 0pt 0pt 20pt},clip,height=.400\textheight,width=.25\textwidth]%
        {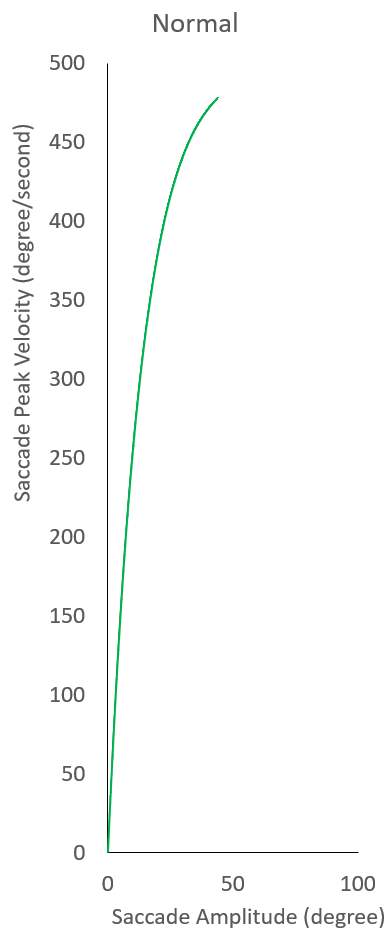}
}
\caption{Main Sequence Relationships 
 \protect\subref{fig:amplitude_vs_velocity_ADHD}
        the relationship between saccade amplitude (degree) and saccade peak velocity (degrees/second) of ADHD subjects,  
        \protect\subref{fig:amplitude_vs_velocity_Non_ADHD}
      the relationship between saccade amplitude (degree) and saccade peak velocity (degrees/second)  of Non-ADHD subjects, and
       \protect\subref{fig:amplitude_vs_velocity_Normal}
       the relationship between saccade amplitude (degree) and saccade peak velocity (degrees/second) of Normal humans.
        }
\label{fig:peak_vel_main_sequence}
\end{figure*}

\begin{figure*}[hbt!]
\centering
\subfigure[Amplitude vs. Duration of ADHD]{%
    \label{fig:amplitude_vs_duration_ADHD}
    \includegraphics[trim={0pt 0pt 0pt 17pt},clip,height=.350\textheight,width=.250\textwidth]%
        {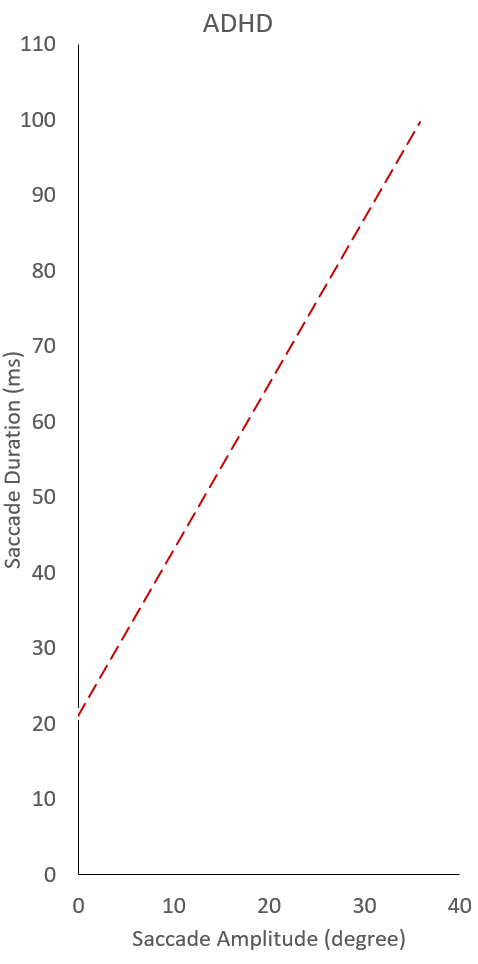}
}
\hspace{1em}%
\subfigure[Amplitude vs. Duration of non-ADHD]{%
    \label{fig:amplitude_vs_duration_Non_ADHD}
    \includegraphics[trim={0pt 0pt 0pt 15.5pt},clip,height=.350\textheight,width=.250\textwidth]%
        {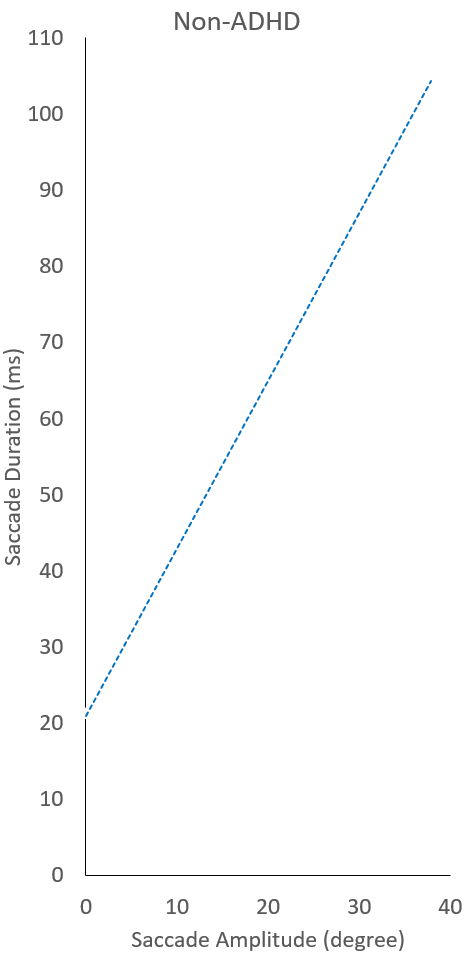}
}
\hspace{1em}%
\subfigure[Amplitude vs. Duration of Normal]{%
    \label{fig:amplitude_vs_duration_Normal}
    \includegraphics[trim={0pt 0pt 0pt 18pt}, clip,height=.350\textheight,width=.250\textwidth]%
        {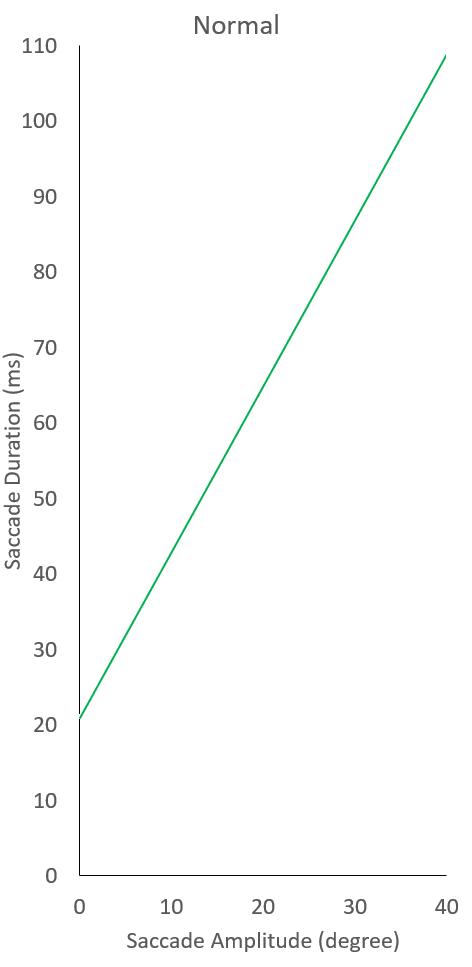}
}
\caption{Main Sequence Relationships,
        \protect\subref{fig:amplitude_vs_duration_ADHD}
        the relationship between saccade amplitude (degree) and duration (ms) of ADHD subjects,  
        \protect\subref{fig:amplitude_vs_duration_Non_ADHD}
      the relationship between saccade amplitude (degree) and duration (ms) of Non-ADHD subjects, and
       \protect\subref{fig:amplitude_vs_duration_Normal}
       the relationship between saccade amplitude (degree) and duration (ms) of Normal humans.
        }
\label{fig:dur_main_sequence}
\end{figure*}

In general, saccades are stereotyped: The relationships between saccade amplitude, saccade peak velocity, and saccade duration are relatively fixed for normal human beings, and are referred to as main sequence relationships.
The two main sequence relationships are:
1) the relationship between saccade amplitude (degree) and duration (ms), and
2) the relationship between saccade amplitude (degree) and saccade peak velocity (degree/second).
We hypothesize that any differences encountered in main sequence relationships could lead to the conclusion that the saccade is not normal.

Figure. \ref{fig:peak_vel_main_sequence} presents the relationships between saccade amplitude and saccade peak velocity in representative ADHD and Non-ADHD adults during the entire session of WMC task.
The data in Figures \ref{fig:amplitude_vs_velocity_ADHD} and \ref{fig:amplitude_vs_velocity_Non_ADHD} show a similar relationship between saccade amplitude and the saccade peak velocity with trend line as well for normal humans (see Figure \ref{fig:amplitude_vs_velocity_Normal}). These results are consistent with the study \cite{fried2014adhd} which describes the main sequence relationship indexing the test of variables of attention.

The data in Figure. \ref{fig:dur_main_sequence} show a similar relationship between saccade amplitude and the saccade duration for ADHD and Non-ADHD adults during the entire WMC task. In addition, it shows similar trend line with normal humans as well (see Figures \ref{fig:amplitude_vs_duration_ADHD}, \ref{fig:amplitude_vs_duration_Non_ADHD}, and \ref{fig:amplitude_vs_duration_Normal}).

\subsection{Machine Learning on Scene-based and Sentence-based Feature Sets}

We obtained all performance metrics using WEKA by executing the selected classifier with a 10-fold cross validation using the both feature sets we developed for the investigation of fixations and saccades within AOIs. The reason for using WEKA is that, it facilitates users to execute machine learning algorithms out-of-the-box and visualize how different algorithms perform for the same data set.

Figure \ref{fig:rspan} presents images of eye gaze patterns from two adults participants, one with and one without ADHD. According to the Figure \ref{fig:rspan}, the adult with ADHD is fixating primarily below the AOIs of stimulus items in sentence including: the words, the decision point, and the item to be remembered (see Figure \ref{fig:aoi}). The adult without ADHD has a larger number of fixations which are in-line with AOIs. 

We selected the same six top performing classifiers listed in Table \ref{tab:classifiers} and we utilize our feature sets which primarily consist of fixation and saccade features within AOIs to train the six classifiers. 

\begin{figure*}[hbt!]
\centering
\subfigure[Amplitude vs. Duration of ADHD]{%
    \label{fig:amp_dur_ADHD_scene}
    \includegraphics[trim={0pt 0pt 0pt 30pt},clip,height=.240\textheight,width=.2\textwidth]%
        {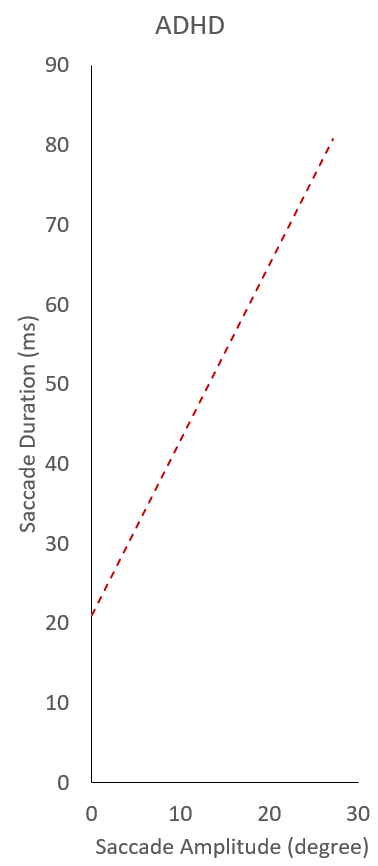}
}
\hspace{1em}%
\subfigure[Amplitude vs. Duration of Non-ADHD]{%
    \label{fig:amp_dur_nonADHD_scene}
    \includegraphics[trim={0pt 0pt 0pt 30pt},clip,height=.240\textheight,width=.2\textwidth]%
        {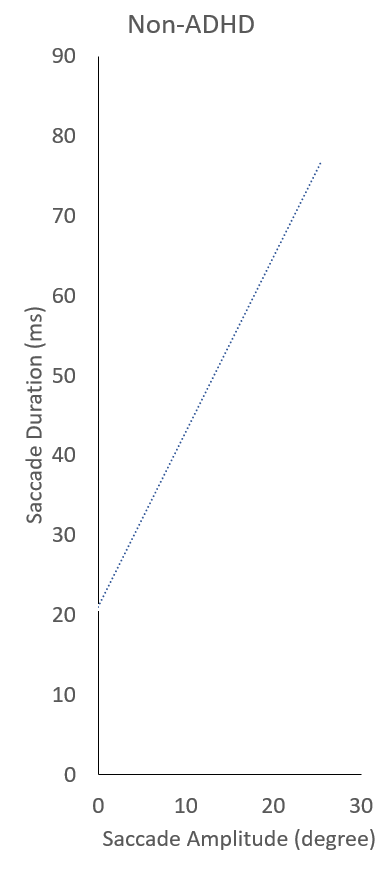}
}
\hspace{1em}%
\subfigure[Amplitude vs. Peak Velocity of ADHD]{%
    \label{fig:amp_vel_ADHD_scene}
    \includegraphics[trim={0pt 0pt 0pt 30pt},clip,height=.240\textheight,width=.2\textwidth]%
        {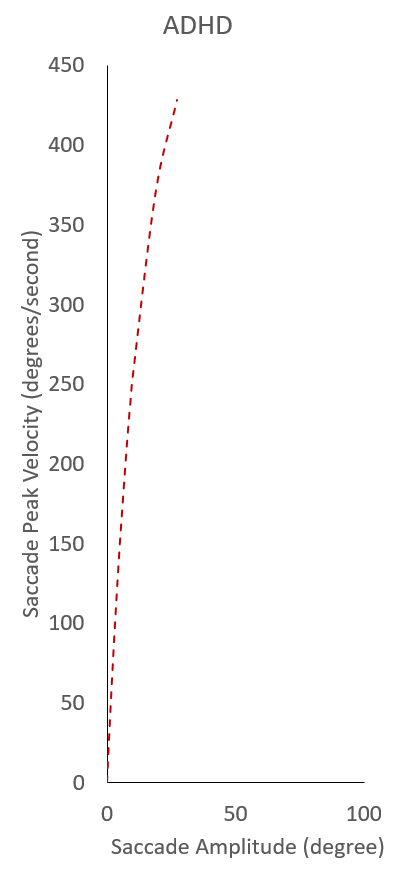}
}
\hspace{1em}%
\subfigure[Amplitude vs. Peak Velocity of Non-ADHD]{%
    \label{fig:amp_vel_nonADHD_scene}
    \includegraphics[trim={0pt 0pt 0pt 30pt},clip,height=.240\textheight,width=.2\textwidth]%
        {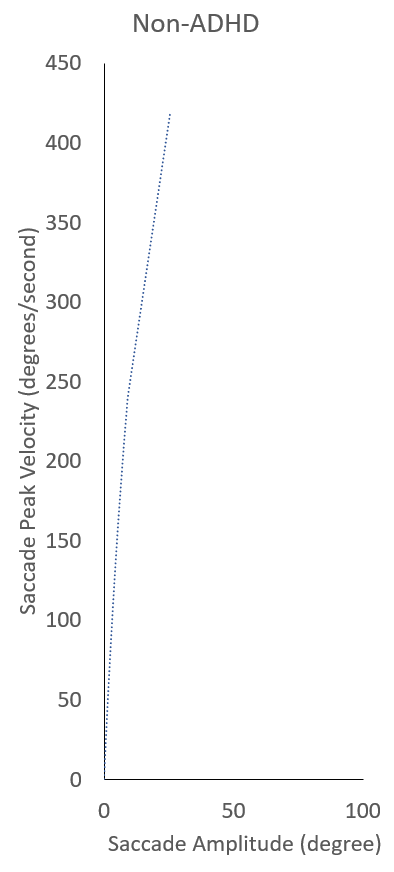}
}
\caption{ Main Sequence relationships obtained from the Scene-based Feature set including eye gaze metrics within the AOIs; \protect\subref{fig:amp_dur_ADHD_scene} Saccade Amplitude vs. Saccade Duration relationship of ADHD participants,  \protect\subref{fig:amp_dur_nonADHD_scene} Saccade Amplitude vs. Saccade Duration relationship of Non-ADHD participants, \protect\subref{fig:amp_vel_ADHD_scene} Saccade Amplitude vs. Saccade Peak Velocity relationship of ADHD participants, and \protect\subref{fig:amp_vel_nonADHD_scene} Saccade Amplitude vs. Saccade Peak Velocity relationship of Non-ADHD participants }
\label{fig:scenes_main_sequence}
\end{figure*}

\begin{figure*}[hbt!]
\centering
\subfigure[Amplitude vs. Duration of ADHD]{%
    \label{fig:amp_dur_ADHD_sentence}
    \includegraphics[trim={0pt 0pt 0pt 30pt},clip,height=.240\textheight,width=.2\textwidth]%
        {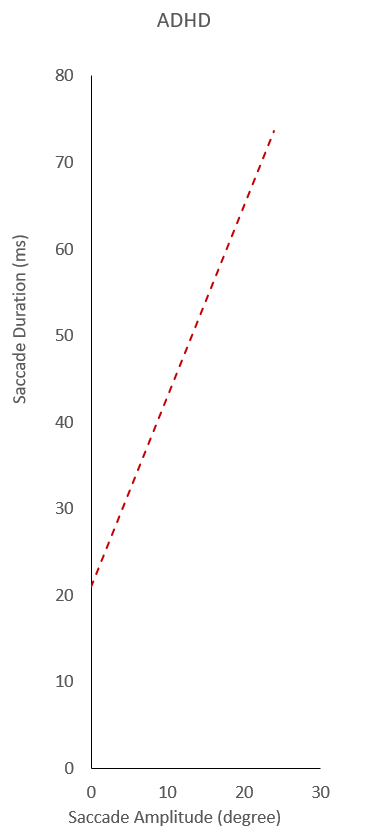}
}
\hspace{1em}%
\subfigure[Amplitude vs. Duration of Non-ADHD]{%
    \label{fig:amp_dur_nonADHD_sentence}
    \includegraphics[trim={0pt 0pt 0pt 30pt},clip,height=.240\textheight,width=.2\textwidth]%
        {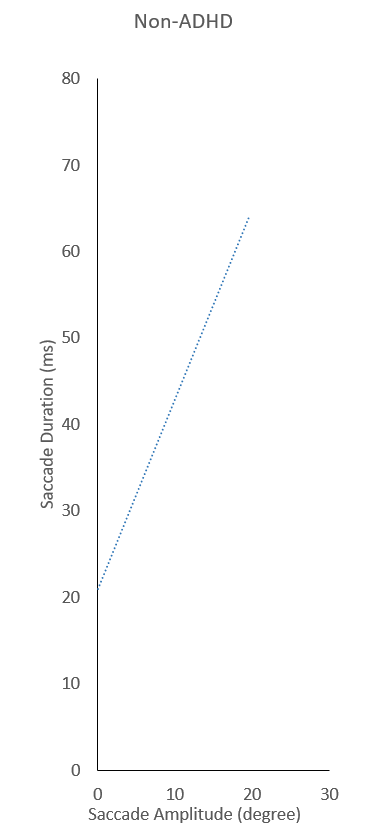}
}
\hspace{1em}%
\subfigure[Amplitude vs. Peak Velocity of ADHD]{%
    \label{fig:amp_vel_ADHD_sentence}
    \includegraphics[trim={0pt 0pt 0pt 30pt},clip,height=.240\textheight,width=.2\textwidth]%
        {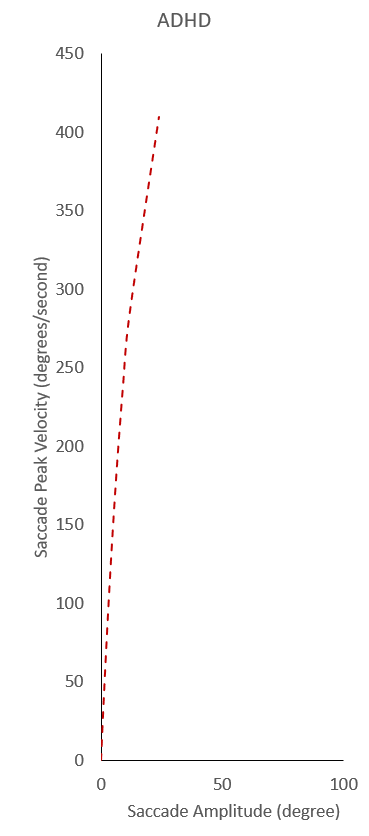}
}
\hspace{1em}%
\subfigure[Amplitude vs. Peak Velocity of Non-ADHD]{%
    \label{fig:amp_vel_nonADHD_sentence}
    \includegraphics[trim={0pt 0pt 0pt 30pt},clip,height=.240\textheight,width=.2\textwidth]%
        {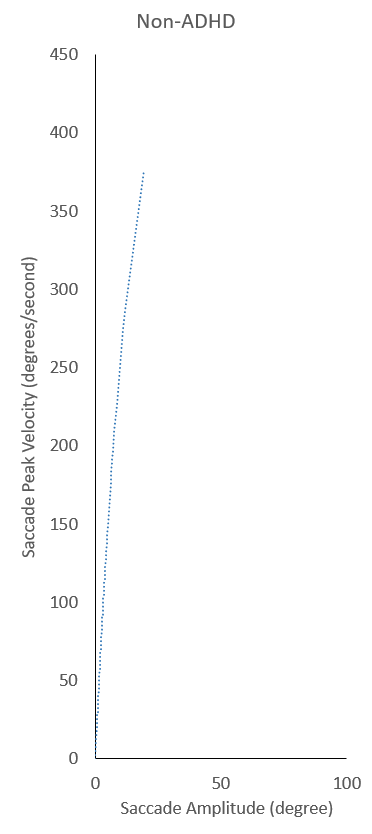}
}
\caption{ Main Sequence relationships obtained from the Sentence-based Feature set including eye gaze metrics within the AOIs ; \protect\subref{fig:amp_dur_ADHD_sentence} Saccade Amplitude vs. Saccade Duration relationship of ADHD participants,  \protect\subref{fig:amp_dur_nonADHD_sentence} Saccade Amplitude vs. Saccade Duration relationship of Non-ADHD participants, \protect\subref{fig:amp_vel_ADHD_sentence} Saccade Amplitude vs. Saccade Peak Velocity relationship of ADHD participants, and \protect\subref{fig:amp_vel_nonADHD_sentence} Saccade Amplitude vs. Saccade Peak Velocity relationship of Non-ADHD participants}
\label{fig:sentences_main_sequence}
\end{figure*}

Table \ref{tab:scenes_fixation_and_saccade_classifiers} lists the classification results of the scene-based feature set. The RandomForest classifier yielded the highest percent accuracy of with 83.33\% indicating that the scene-based feature set alone classifies a diagnosis of ADHD with greater than 80\% accuracy. The Kstar classifier yielded the lowest percent accuracy at 63.04\% for the scene-based feature set.

Table \ref{tab:sentences_fixation_and_saccade_classifiers} lists the results of the sentence-based feature set. The RandomForest classifier yielded the highest percent accuracy of with 86.20\% indicating that the sentence-based feature set classifies a diagnosis of ADHD with greater than 85\% accuracy. For sentence-based feature set, K star classifier yielded the lowest percent accuracy at 71.83\%. 

Since any differences encountered in main sequence relationships could lead to the conclusion that the saccade is not normal, we plot Figure. \ref{fig:scenes_main_sequence} and  \ref{fig:sentences_main_sequence} to analyze the main sequence relationships of the two feature sets we generated using AOIs. Figure. \ref{fig:amp_vel_ADHD_scene} presents the relationship between saccade amplitude and the saccade peak velocity in representative ADHD adults and \ref{fig:amp_vel_nonADHD_scene} presents the relationship between saccade amplitude and the saccade peak velocity in representative Non-ADHD adults when using the scene-based feature set during the entire WMC task. The data in Figures \ref{fig:amp_vel_ADHD_scene} and \ref{fig:amp_vel_nonADHD_scene} show a similar relationship between saccade amplitude vs. the saccade peak velocity and similar saccade amplitude range for both ADHD and Non-ADHD subject groups. The data in Figure \ref{fig:amp_dur_ADHD_scene} and \ref{fig:amp_dur_nonADHD_scene} show a similar relationship between saccade amplitude and the saccade duration for ADHD and Non-ADHD adults when using the scene-based feature set during the entire WMC task. The results indicate when considering AOIs of scence-based sentences, ADHD and Non-ADHD adults display similarities in main sequence relationships which are similar to normal humans as well.

The data in Figures \ref{fig:amp_vel_ADHD_sentence} and \ref{fig:amp_vel_nonADHD_sentence} show a similar relationship between saccade amplitude vs. the saccade peak velocity for both ADHD and Non-ADHD subject groups when using the sentence-based feature set during the entire WMC task. The data in Figure \ref{fig:amp_dur_ADHD_sentence} and \ref{fig:amp_dur_nonADHD_sentence} also show a similar relationship between saccade amplitude and the saccade duration for ADHD and Non-ADHD adults. These relationships are similar to the relationships obtained from the scene-based feature set (see Figure \ref{fig:scenes_main_sequence}).

\section{Discussion}
Since participants are presented with varying sets of 2-5 sentences, we developed one feature set considering the AOIs in the first sentence of all the sentence sets (scene-based feature set) and the other feature set considering the AOIs of all the 42 sentences in the RSPAN task (sentence-based feature set). We used common AOIs among all the participants, thus they are static shapes and would not change from one subject to another. 
In the case of static AOIs, the granularity of the sentence-based feature set is increased when compared to the granularity of the scene-based feature set. As a result, we observe better accuracy in each classifier(See Table ~\ref{tab:sentences_fixation_and_saccade_classifiers}).

\begin{table}[hbt!]
\centering
  \caption{Classification of Eye Fixation and Saccade Features within AOIs of Scence-based during WMC}
  \label{tab:scenes_fixation_and_saccade_classifiers}
  \begin{tabular}{ccccl}
    \toprule
    Classifier & Precision & Recall & F1 & Accuracy\\
    \midrule
    J48 & 0.75 & 0.75 & 0.75 & 75.36 \\
    LMT & 0.79 & 0.79 & 0.79 & \textbf{79.71} \\
    RandomForest & 0.83 & 0.83 & 0.83 & \textbf{83.33} \\
    REPTree & 0.70 & 0.70 & 0.69 & 70.29 \\
    K* & 0.63 & 0.63 & 0.63 & 63.04 \\
    Bagging & 0.80 & 0.79 & 0.79 & \textbf{79.71} \\
  \bottomrule
\end{tabular}
\end{table}

\begin{table}[hbt!]
\centering
  \caption{Classification of Eye Fixation and Saccade Features within AOIs of Sentence-based  during WMC}
  \label{tab:sentences_fixation_and_saccade_classifiers}
  \begin{tabular}{ccccl}
    \toprule
    Classifier & Precision & Recall & F1 & Accuracy\\
    \midrule
    J48 & 0.79 & 0.79 & 0.79 & 79.39 \\
    LMT & 0.82 & 0.82 & 0.82 & \textbf{82.61} \\
    RandomForest & 0.86 & 0.86 & 0.86 & \textbf{86.20} \\
    REPTree & 0.82 & 0.82 & 0.82 & 82.04 \\
    K* & 0.72 & 0.71 & 0.71 & 71.83 \\
    Bagging & 0.84 & 0.83 & 0.83 & \textbf{83.93} \\
  \bottomrule
\end{tabular}
\end{table}

Consideration of fixation as well as saccade features set according to stimulus AOIs, lead us to classify a diagnosis of ADHD with greater than 80\% accuracy. Our results confirm the utility of eye movement feature set generated according to stimulus AOIs indexing WMC as a predictor of a diagnosis of ADHD in adults. RandomForest classifiers performed best in-terms of predicting  a classification of ADHD with 86.20\% percent accuracy by using sentence-based feature set representing a physiological measure of visual attention during a WMC task. 

Since we are utilizing the eye gaze metrics calculated by Tobii Studio analysis software within the manually marked static AOI boundaries for the development of our feature sets, there might be instances where we have less data points. The static AOIs may not be enough in terms of the area boundary to capture eye gaze metrics of some of the participants. We did not consider device error or human error when creating the AOI boundaries. In the future, we are interested in identifying AOIs dynamically for each participant in each sentence.

\section{Conclusions}

The purpose of this feasibility study was to determine if patterns of saccades and stability of fixations generated when completing a measure of WMC, the RSPAN task, would create a feature set which could be used to differentiate a diagnosis of ADHD for adults. We used accuracy, precision, recall, and f-measure as the evaluation metrics. While fixation features, saccade features, and a combination of saccade and fixtaion features accurately predicted the classification of ADHD with an accuracy of greater than 78\%, saccade features were the best predictors with an accuracy of 91\%. The results of this feasibility study confirmed the utility of a combination of fixation, and saccade feature set generated within AOIs while completing RSPAN tasks as a predictor of a diagnosis of ADHD in adults. Tree-based classifiers performed best in-terms of predicting a classification of ADHD with 86\% percent accuracy using physiological measures of sustained visual attention within AOIs during a WMC task.

These results are consistent with previous studies confirming significant differences in saccadic behaviors for people with ADHD \cite{lee2015saccadic,roberts2011separating}. During the RSPAN task, the rapid movement of the eye across the scan path from one fixation point to the other yields a more accurate classification of ADHD than the ability to sustain gaze. These preliminary results indicate that detailed and discrete eye gaze metrics during a measure of attention control (i.e. WMC) provide unique indices of ADHD and offer physiological insight regarding cognitive resources underlying WMC, an important cognitive construct responsible for behavioral inhibition and attention monitoring. Moreover, they are consistent with previous investigations finding that adults with ADHD demonstrate similar broad visual attention patterns as adults without ADHD but different scan patterns \cite{krejtz2015gaze} and different pupillometry metrics as a function of visual cue type \cite{michalek116pupil}.

\section{Future Directions}
The results of this feasibility study confirm the utility of eye movement feature set indexing WMC as a predictor of a diagnosis of ADHD in adults. RandomForest classifiers performed best in-terms of predicting a classification of ADHD with 91.14\% percent accuracy by combining saccade feature set representing a physiological measure of visual attention during a WMC task. Tree-based classifiers performed best in-terms of predicting a classification of ADHD with 86\% percent accuracy using eye gaze metrics within AOIs during a WMC task. This project is a necessary first step in delineating a feature set of eye gaze metrics captured within AOIs which represent physiological diagnostic criteria, including executive attention in adult ADHD.

In the future, we will expand the experimental studies to further analyze eye gaze metrics according to dynamically changing stimulus AOIs with respect to the participants using a larger sample size. Specifically, creating a boundary for the AOIs; the sentence, the word which determine sentence accuracy, the visual point of decision, and the item to be remembered. Identifying these AOIs dynamically for each participant will enable us to generate a detailed feature set which could be utilized to classify a diagnosis of ADHD with a greater percentage of accuracy than of this study.

\newpage
\bibliographystyle{abbrv}

\end{document}